\documentclass[twocolumn]{aastex63}

\usepackage{newtxmath}
\usepackage{graphicx}
\usepackage{xcolor}
\usepackage[caption=false]{subfig}

\shorttitle{Probing the local ISM with PSR\,B1133$+$16 scintillometry}
\shortauthors{McKee, J.\,W., Zhu, H., Stinebring, D.\,R., \& Cordes, J.\,M.}
\graphicspath{{./}{}}

\begin{document}

\title{Probing the local interstellar medium with scintillometry of the bright pulsar B1133+16}

\correspondingauthor{James W. McKee}
\email{jmckee@cita.utoronto.ca}

\author[0000-0002-2885-8485]{James W. McKee}
\affiliation{Canadian Institute for Theoretical Astrophysics, University of Toronto, 60 Saint George Street, Toronto, ON M5S 3H8, Canada}

\author[0000-0001-9027-4184]{Hengrui Zhu}
\affiliation{Department of Physics \& Astronomy, Oberlin College, Oberlin, OH 44074}

\author[0000-0002-1797-3277]{Daniel R. Stinebring}
\affiliation{Department of Physics \& Astronomy, Oberlin College, Oberlin, OH 44074}

\author[0000-0002-4049-1882]{James M. Cordes}
\affiliation{Cornell Center for Astrophysics and Planetary Science, and Department of Astronomy, Cornell University, Ithaca, NY 14853, USA}




\begin{abstract}

The interstellar medium hosts a population of scattering screens, most of unknown origin. Scintillation studies of pulsars provides a sensitive tool for resolving these scattering screens and a means of measuring their properties. In this paper, we report our analysis of 34 years of Arecibo observations of PSR\,B1133+16, from which we have obtained high-quality dynamic spectra and their associated scintillation arcs, arising from the scattering screens located along the line of sight to the pulsar. We have identified six individual scattering screens that are responsible for the observed scintillation arcs, which persist for decades. 
Using the assumption that the scattering screens have not changed significantly in this time, we have modeled the variations in arc curvature throughout the Earth's orbit and extracted information about 
the placement, orientation, and velocity of five of the six screens, with the highest-precision distance measurement placing a screen at just $5.46^{+0.54}_{-0.59}$\,pc from the Earth. We associate the more distant of these screens with an under-dense region of the Local Bubble.

\end{abstract}


\ \


\section{Introduction}\label{sec:intro}
The propagation of pulses from radio pulsars is affected by temporal and spatial variations in integrated electron column number density and by small-scale inhomogeneities in the ionized plasma of the interstellar medium (ISM, e.g. \citealp{ars95}). All ISM effects scale rapidly with observing frequency as $\nu^{-\gamma}$ (with $\gamma=2$ and $\gamma\sim4$ for dispersion and scattering/scintillation
respectively, e.g. \citealp{cwb85}, \citealp{gs06b}), and radio astronomy of pulsars has allowed for the number density of the ionized plasma of the ISM to be directly probed and turbulence to be inferred along different lines of sight (e.g. \citealp{tmc+21}, \citealp{aab+21}, \citealp{lmj+16}, \citealp{acr81}, \citealp{ars95}, \citealp{rcb+20}).

Time-varying inhomogeneities in the line of sight electron column give rise to two important effects: scattering and scintillation. Multi-path scattering causes temporal broadening of the pulses and angular broadening of the image, analogous to a convolution of the intrinsic pulse shape with an asymmetric broadening function (e.g. described in detail in \citealp{wil72}, \citealp{wil73}, \citealp{wil74}, and see recent work by \citealp{gk16}, \citealp{mls+18}), which is canonically assumed to be exponential, although a Kolmogorov medium with a small inner scale will depart from this (see e.g. \citealp{lj75}, \citealp{lr00}).

A phenomenon related to scattering is scintillation; since the pulsar has an angular size that is small enough for it to be considered a point source, the varying inhomogeneities along the line of sight cause the scattered rays to interfere with each other constructively and destructively, giving rise to an interference pattern.
As the Earth moves across this interference pattern, the observed flux of the pulsar is seen to vary as a function of time and observing frequency.
Studies of the scintillation properties of pulsars have proved to be a very sensitive tool for probing the ISM, studying binary pulsars (\citealp{lyn84}, and see recent work by \citealp{rcb+20}), and identifying structures within the ISM. \cite{smc+01} were the first to demonstrate that the two-dimensional Fourier transform of a dynamic spectrum produces scintillation arcs in the `secondary spectrum' which are interpreted as arising from discrete scattering screens along the line of sight to the pulsar. The curvatures\footnote{Strictly, this quantity is the `quadratic constant of proportionality', although we use the term `curvature' in keeping with the commonly-used nomenclature in pulsar scintillometry.} of these arcs are highly dependent on the velocity as well as the geometry of scattering screens, and studying their variation with time has allowed very precise measurements of scattering regions (e.g. \citealp{sim19}, \citealp{rcb+20}).

The origin of these scattering screens has been variously attributed to known sources including: circumstellar structures in the vicinity of hot stars \citep{wtb+17}, H\,II regions along the line of sight to the pulsar \citep{ssh+00}, and expanding supernova remnants (e.g. \citealp{lr00}). However, it has not always been possible to associate pulsars with any known structures making the origin of some scattering screens mysterious. 
In these cases, the inhomogeneities have instead been explained by: structures and boundaries within the Local Bubble \citep{bgr98}, corrugated current sheets, coincidentally angled close to the line of sight to the pulsar \citep{pl14}, and thin but highly-elongated lenses which lead to an apparent over-density due to their projection along the line of sight \citep{rbc87}.
Such explanations are particularly necessary in the cases where multiple scattering screens are identified within a small region between the Earth and the pulsar (see \citealp{ps06a}).
Observations of several pulsars have shown evidence of highly anisotropic scattering, where the scattered images of the pulsars are highly elongated along particular axes (\citealp{wms+04}, \citealp{bmg+10}, \citealp{rszm21}, \citealp{sro19}). Such anisotropies in scattering may result from anisotropic plasma structure in the ISM e.g. corrugated plasma sheets viewed from a grazing angle \citep{pl14}, or `noodles' of plasma which are stabilized by magnetic re-connection \citep{gwinn19}.

A notable pulsar where multiple scintillation arcs have been detected is PSR\,B1133+16 (known as J1136+1551 in the J2000 equinox, and see \citealp{crsc06}, \citealp{ps06a}). We summarize some of this pulsar's important astrometric parameters in Table \ref{tab:b1133_values}. This is a nearby pulsar ($372\pm3$\,pc, 1-$\sigma$ uncertainties are used throughout), with a very high transverse velocity of $659.7^{+4.2}_{-4.5}$\,km\,s$^{-1}$, where both quantities have been measured using very long baseline interferometry (VLBI, \citealp{dgb+19}\footnote{The reference lists an uncertainty of zero for the distance, so we have calculated the distance from the more precise parallax values listed on the PSRPI website: https://safe.nrao.edu/vlba/psrpi/home.html}). This combination of the pulsar's close distance and high speed leads to rapid variations in its observed scintillation, which has been studied extensively (e.g. \citealp{tr07}). PSR\,B1133+16 is one of the brightest-known pulsars ($20\pm10$\,mJy at 1382\,MHz, \citealp{jvk+18}), which makes it an excellent target for studying the low signal-to-noise (S/N) secondary spectra parabolas. PSR\,B1133+16 is one of a few pulsars for which the distances to the scattering screens have been estimated \citep{ps06a}. We note, however, that the distances to the PSR\,B1133+16 screens reported by \cite{ps06a} assumes a screen orientation such that the major axis of the anisotropic image is aligned with the declination axis (see Section \ref{sec:annual_variations})
and therefore represent a lower distance limit. 
Precise estimation of the screen distance requires this angle to be known, either through direct VLBI measurements as in \cite{bmg+10}, or by modeling the contribution from the Earth's orbit, which we present in detail in Section \ref{sec:annual_variations}.

\begin{table}
\caption{Summary of astrometric parameters for PSR\,B1133+16. All parameters are measured using VLBI, and are presented in \cite{dgb+19}. Note that this reference lists an uncertainty of zero for the pulsar distance, and we have instead calculated the distance from the more precise parallax values listed on the PSRPI website.}
\label{tab:b1133_values}
\centering 
\begin{tabular} {c c}
\hline
\hline
\multicolumn{2}{c}{Astrometric parameters}\\
\hline
Right ascension, $\alpha$ (h:m:s) \dotfill & 11:36:03.1198(1) \\
Declination, $\delta$ (d:m:s) \dotfill & 15:51:14.183(1) \\
Ecliptic longitude, $\lambda$ ($^{\circ}$) \dotfill & 168.15 \\
Ecliptic latitude, $\beta$ ($^{\circ}$) \dotfill & 12.17 \\
Galactic longitude, $l$ ($^{\circ}$) \dotfill & 241.895 \\
Galactic latitude, $b$ ($^{\circ}$) \dotfill & 69.196 \\
Galactic height, $z$ (pc) \dotfill & 346 \\
\vspace{0.15cm}
Proper motion in $\alpha$, $\mu_{\alpha}$ (mas\,yr$^{-1})$ \dotfill & $-73.785^{+0.031}_{-0.010}$ \\
Proper motion in $\delta$, $\mu_{\delta}$ (mas\,yr$^{-1})$ \dotfill & $366.569^{+0.072}_{-0.055}$ \\
Distance (pc) \dotfill & $372\pm3$ \\
Transverse velocity (km\,s$^{-1}$) \dotfill & $659.7^{+4.2}_{-4.5}$ \\
\vspace{-0.3cm}
 & \\
\hline
\hline
\end{tabular}
\end{table}

The remainder of the paper has the following structure: in Section \ref{sec:theory}, we outline the relevant theory for describing scintillation variations and their applications in isolating scattering screen properties. In Section \ref{sec:obsanddata}, we describe our observations and data reduction techniques. We present the results of our analysis in Section \ref{sec:results} and discuss their implications of our findings in Section \ref{sec:discussion}. Finally, we make concluding remarks in Section \ref{sec:conclusions}.
\begin{figure*}[!htbp]
    \vspace{0.4cm}
    \centering
	\includegraphics[width=\textwidth]{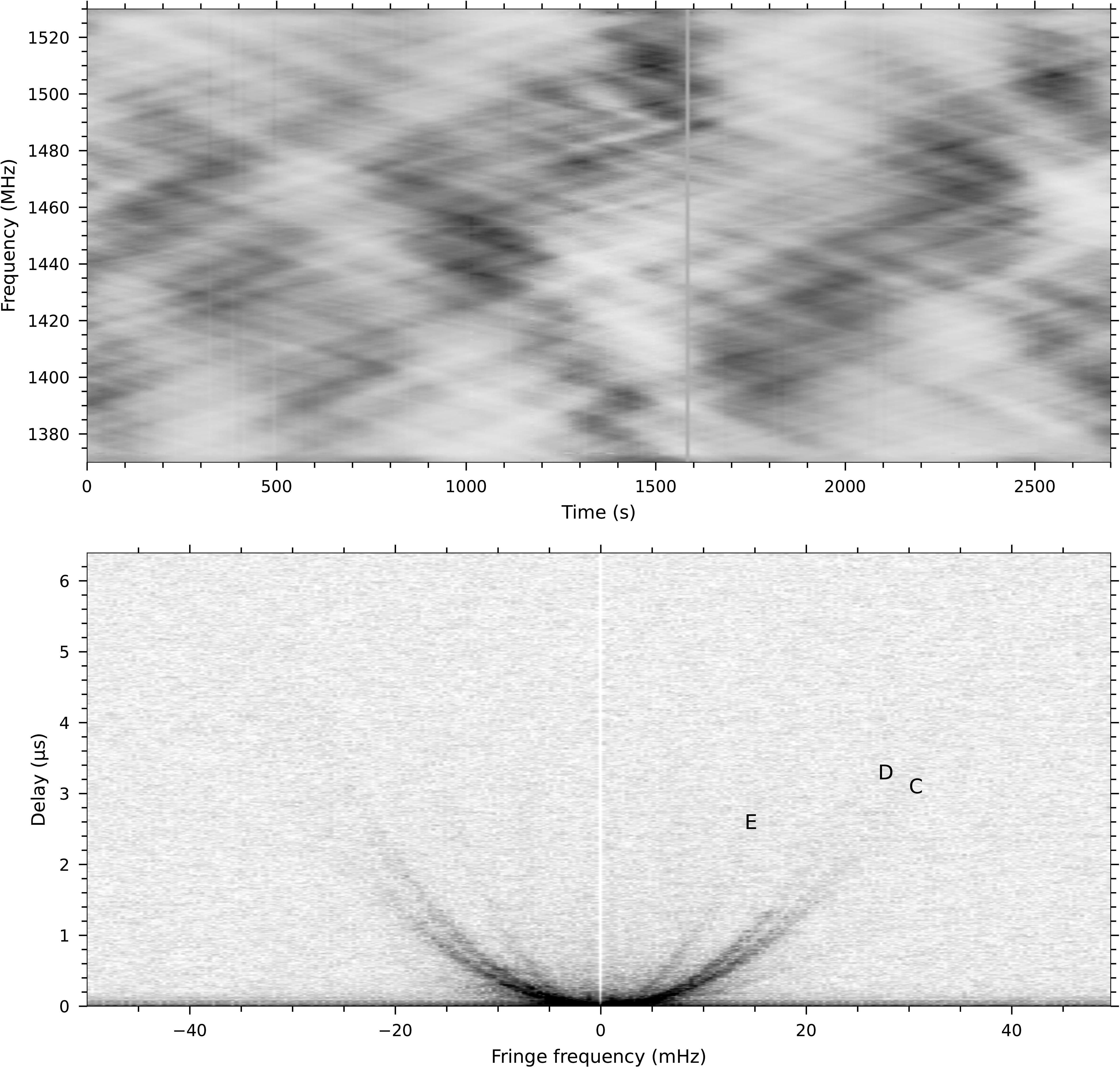}	
	\caption{An example of a PSR\,B1133+16 dynamic spectrum (\textit{above}, linear grayscale), taken at a center frequency of 1450\,MHz. The observation was made on MJD\,57098 (day number 119 in Figure \ref{fig:curvature_annual}), with darker colors corresponding to greater flux density. One sub-integration containing RFI has been removed. The associated secondary spectrum (\textit{below}, using a logarithmic grayscale, and with the center fringe frequency removed), the squared modulus of a two-dimensional Fourier transform of the dynamic spectrum, displays at least three separate arcs (labeled C, D, E, in order of increasing curvature), for which we have measured curvatures of 0.0041(5), 0.0054(3), and 0.013(2)\,s$^{3}$ (scaled to a frequency of 1400\,MHz). The arcs correspond to persistent discrete scattering screens, which we observe throughout the length of our data set. The sharply-defined arcs indicate the presence of weak scintillation in this observation. Note that there is a hint of a lower-curvature arc which is consistent with the Arc B curvature, but the S/N is too low for the curvature to be precisely measured.}
	\label{fig:B1133_dynspec}
\end{figure*}
\section{Scintillation theory}\label{sec:theory}
We follow the theoretical treatment of thin screen scattering as presented in e.g. \cite{wms+04} and \cite{crsc06}. The pulsed flux from pulsars is affected by interstellar scintillation such that the observed intensity $I$ varies as a function of time $t$ and observing frequency $\nu$, i.e. $I(\nu,t)$, known as the dynamic spectrum. The two-dimensional Fourier transform (represented by a tilde) recasts variations in observed intensity in terms of time delay $\tau$ and Doppler frequency $f_{\text{D}}$, and we sum over pairs of interfering images (represented by $j,k$) with magnification $\mu$ relative to the un-lensed image:
\begin{equation}\label{eqn:intensity}
    \left\vert\widetilde{I}(\tau,f_{\text{D}})\right\vert^2 \propto \sum_{j,k}\mu_{j}\mu_{k}\delta(f_{\text{D}}-f_{\text{D}j,k})\delta(\tau-\tau_{j,k}).
\end{equation}
The above equation defines a secondary spectrum, the two-dimensional power spectrum of the dynamic spectrum.
We characterize scattering screens by their effective distance and velocity vector $D_{\text{eff}}$ and $\bf{V}_{\text{eff}}$ respectively. 
The effective distance of the scattering screen $D_{\text{eff}}$ is defined in terms of its fractional distance $s$ from the pulsar to the observer (i.e. a small value of $s$ indicates that a scattering screen is close to the pulsar), 
\begin{equation}\label{eqn:d_eff}
    D_{\text{eff}}=d_{\text{psr}}\frac{1-s}{s} ,
\end{equation}
where
\begin{equation}\label{eqn:fractional_dist}
    s=1-\frac{d_{\text{screen}}}{d_{\text{psr}}} ,
\end{equation}
i.e. $0<s<1$. 
The effective velocity of the interference pattern in the plane of the observer is a combination of the pulsar, Earth, and scattering screen velocities (\citealp{cr98}, \citealp{grl94}):
\begin{equation}\label{eqn:V_eff1}
    \textbf{V}_{\text{eff}}=-\textbf{V}_{\text{psr}}\frac{1-s}{s}+\frac{\textbf{V}_{\text{screen}}}{s}-\textbf{V}_{\text{Earth}} ,
\end{equation}
where all of the velocities above refer to the components perpendicular to the line-of-sight (as opposed to the full three-dimensional velocities). When the scattering screen is highly anisotropic i.e. when it composes of a group of parallel and highly elongated plasma lenses, the resulting image of the pulsar is also highly elongated, with its major axis perpendicular to the lenses on the scattering screen. In this scenario, only the effective velocity parallel to the image can be extracted (see Equation \ref{eqn:f_d}).

The effective distance and velocity allows us to define the axes of the secondary spectrum in  terms of the temporal and spectral delays, given by
\begin{equation}\label{eqn:tau}
    \tau=\frac{D_{\text{eff}}(\theta^{2}_{j}-\theta^{2}_{k})}{2c} 
\end{equation}
and
\begin{equation}\label{eqn:f_d}
    f_{\text{D}}=\frac{\mathbf{V}_{\text{eff}}\cdot\left(\boldsymbol{\theta}_{j}-\boldsymbol{\theta}_{k}\right)}{\lambda} 
\end{equation}
respectively, where $\lambda$ is the observing wavelength, and $\theta$ describes the observed angle of a scattered image.
In highly anisotropic scattering, the interference between the scattered images and the line-of-sight image of the pulsar leads to parabolic arcs in the secondary spectrum described by $\tau = \eta f_{\text{D}}^{2}$ \citep{smc+01}, with curvatures given by
\begin{equation}\label{eqn:eta}
    \eta=\frac{\lambda^{2}}{2c}\frac{D_{\text{eff}}}{V_{\alpha}^{2}} ,
\end{equation}
where $V_{\alpha}=|\textbf{V}_{\text{eff}}|\cos\alpha$, and $\alpha$ is the angle between the effective velocity and the major axis of the image.
The parabola curvature is seen to vary over time due to the proper motion of the pulsar, the Earth's orbital motion, and the movement of the scattering screen (and the pulsar orbit, when in a binary system).

\begin{table}
\caption{Summary of the observational configurations used in our data set and the number of observations taken with each, with the historical (pre-2015) parameters listed separately from those of the 2015 campaign.}
\label{tab:observationsummary}
\centering 
\begin{tabular} {c c c}
\hline
\hline
Center frequency (MHz) & Bandwidth (MHz) & $N_{\text{obs}}$ \\
\hline
\hline
\multicolumn{3}{c}{MJD\,44563 -- 53190}\\
\hline
321 -- 333 & 3.125 -- 25.000 & 10 \\
430 & 10 & 30 \\
800 -- 820 & 40 & 3 \\
1175 & 100 & 13 \\
1525 & 50 & 1 \\
2150 -- 2250 & 100 & 2 \\
\hline
\multicolumn{3}{c}{MJD\,57058 -- 57201}\\
\hline
327 & 53.333 & 11 \\
432 & 26.667 & 18 \\
1450 & 160 & 8 \\
\hline
\end{tabular}
\end{table}

We note that although 
the component of the pulsar velocity parallel to the line of sight
does not contribute to the scintillation time it does, of course, change the distance to the pulsar. In the case of PSR\,B1133+16, this change is negligible: a parallel velocity of $\sim100$\,km\,s$^{-1}$ leads to a change in pulsar distance of $\sim0.0035$\,pc ($\sim700$\,au) over our 34-yr data set, and we use the assumption that the screen properties have not changed significantly in that time. However, in the case where the pulsar is very fast-moving, the length of the data set is very long, and a screen is very close to the pulsar, the parallel component of the velocity becomes significant. A pulsar moving relative to a fixed screen experiences a change in screen placement over time of
\begin{equation}\label{eqn:delta_S_fixed_screen}
    \Delta s(t) = \frac{d_{\text{screen}}V_{\parallel}t}{d_{\text{psr}}^{0}\left(d_{\text{psr}}^{0}+V_{\parallel}t\right)}  ,
\end{equation}
for a pulsar at an initial distance $d_{\text{psr}}^{0}$ with a parallel velocity component of $V_{\parallel}$.
In the case where the distance between the screen and the moving pulsar remains constant (e.g. the bow shock nebula associated with PSR\,J0437$-$4715, where the screen moves with the pulsar, \citealp{bbb93}), the change in screen placement over time becomes
\begin{equation}\label{eqn:delta_S_moving_screen}
    \Delta s(t) = \frac{\left(d_{\text{screen}}^{0}-d_{\text{psr}}^{0}\right)V_{\parallel}t}{d_{\text{psr}}^{0}\left(d_{\text{psr}}^{0}+V_{\parallel}t\right)}  ,
\end{equation}
when the screen has an initial distance $d_{\text{screen}}^{0}$. This assumes that other properties of the screen remain constant, which may not be the case. For example, significant differences in the distance of the shock contour in the Guitar Nebula are observed, due to changes in the ambient density (\citealp{cc04}, \citealp{occd21}), which would further complicate modeling of the screen.

\begin{figure}
    \vspace{0.5cm}
	\includegraphics[width=\columnwidth]{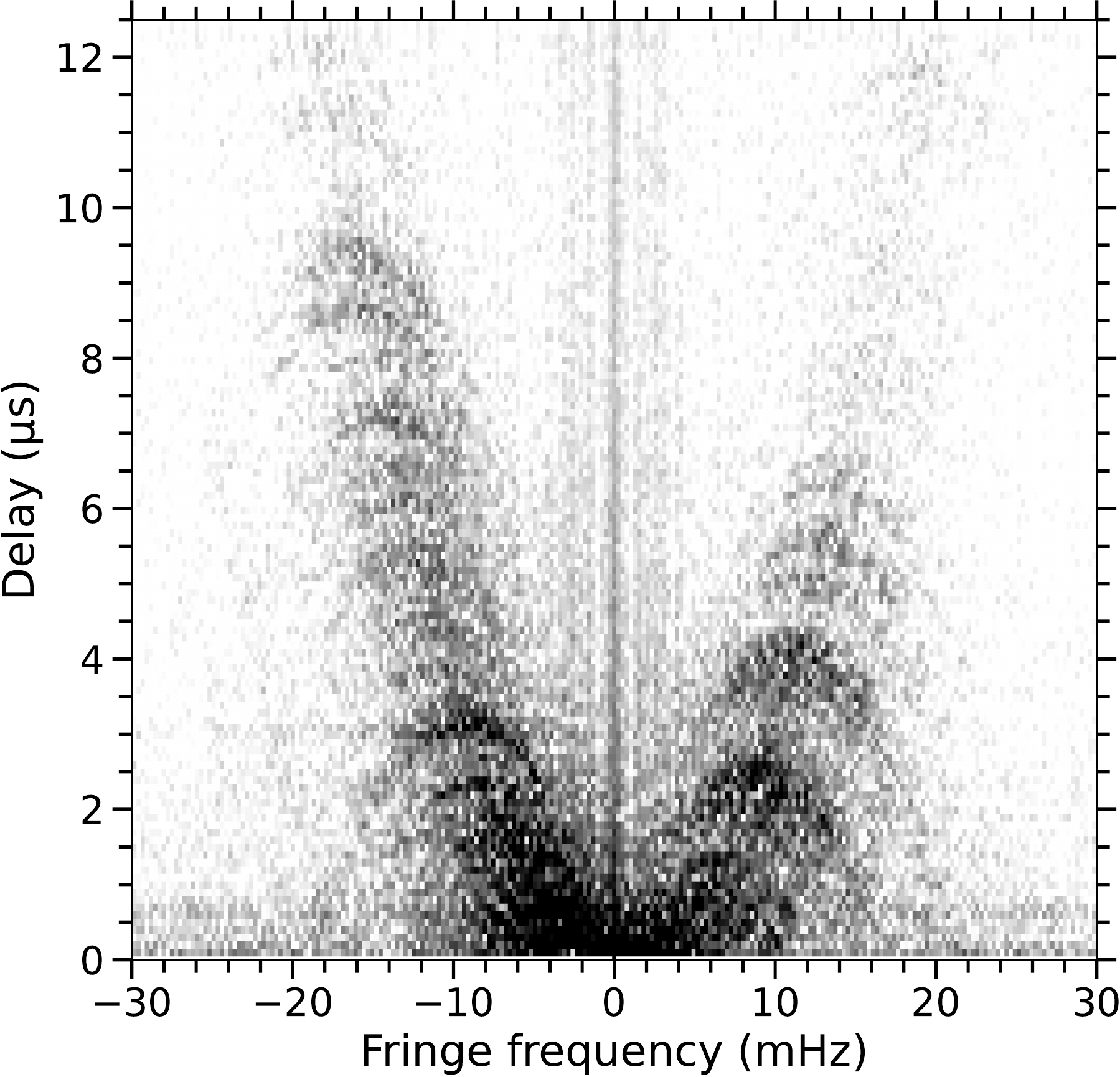}	
	\centering
	\caption{Inverted arclets detected in the secondary spectrum of an observation taken on MJD\,45988 (day 332 in Figure \ref{fig:curvature_annual}), at a center frequency of 430\,MHz.}
	\label{fig:inverted_arclets}
\end{figure}

\section{Observations and Data Reduction}\label{sec:obsanddata}
Our data set is composed of observations taken over a 34-yr period using the Arecibo telescope. Between 1980\,--\,2015, observations were taken sporadically, with gaps of many years in the data set (the longest being $\sim10$\,yr). Observations were taken using the Mock spectrometer backend\footnote{https://www.naic.edu/ao/scientist-user-portal/astronomy/backends}, at Nyquist sampling rates of typically $6.4$\,--\,$38.4\,\muup$s, and with data recorded using 10 second sub-integrations. Using the approach outlined in \cite{hsb+03}, spectra were determined by recording the flux densities ($S$) of the on-pulse and off-pulse phase regions in each frequency channel ($\nu$) and in each sub-integration ($t_{i})$, and weighting by the off-pulse average from the entire observation i.e.
\begin{equation}\label{eqn:dynspecflux}
S(t_{i},\nu)=\frac{S(\nu)^{\text{on}}_{i}-S(\nu)^{\text{off}}_{i}}{\langle S(\nu)^{\text{off}} \rangle}  .
\end{equation}
\begin{figure*}[!htbp]
    \vspace{0.5cm}
	\includegraphics[width=\textwidth]{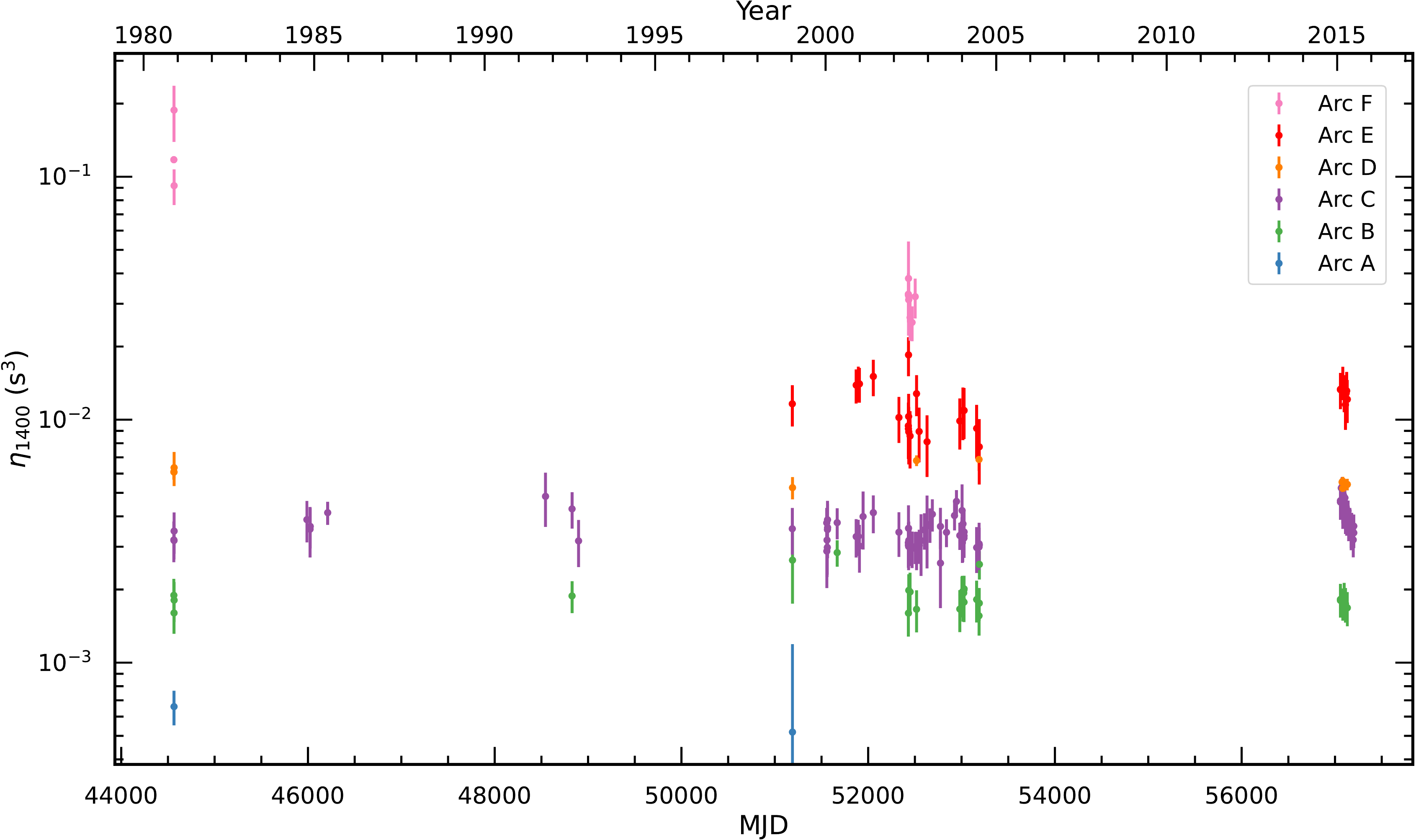}	
	\centering
	\caption{Variations in arc curvature $\eta$ as a function of time (scaled to a reference frequency of 1400\,MHz), for our entire data set. Multiple arcs are resolvable, corresponding to specific ranges of $\eta$ which we attribute to specific scattering screens, and are color coded. The detection of these arcs over a long base line demonstrates that the underlying screens persist for decades, over the length of our entire data set.}
	\label{fig:curvature_vals}
\end{figure*}

In addition to the historical data, our data set includes observations taken as part of a dense campaign, carried out between January and July 2015. During this time, 21 multi-frequency observations were taken with an approximately weekly cadence. Observations were typically 45 minutes in length, primarily at a center frequency of 432\,MHz with 26.7\,MHz of bandwidth. Immediately after most of these observations, the pulsar was observed again, using a different frequency: either 327\,MHz with 53.3\,MHz of bandwidth, or 1450\,MHz (160\,MHz bandwidth). The number of spectral channels used in the 327-MHz, 430-MHz, and 1450-MHz observations were, respectively, 4096, 2048, and 2048, resulting in a Nyquist sampling rate of 38.4\,$\muup$s for the two lower frequencies, and 6.4\,$\muup$s for 1450\,MHz.
Our observational configurations are summarized in Table \ref{tab:observationsummary}.

The high sensitivity of the Arecibo telescope, together with the fine spectral resolution of our data set, has enabled high-quality dynamic spectra to be generated from all of our observations. An example of a dynamic spectrum from an observation at 1450\,MHz is shown in the upper panel of Figure \ref{fig:B1133_dynspec}. 
All of our dynamic spectra were converted to secondary spectra 
as the squared modulus of the
two-dimensional Fourier transform (described in Section \ref{sec:theory}).

\section{Results and analysis}\label{sec:results}
We observe scintillation arcs in the secondary spectra from all of our observations and see that many observations show multiple simultaneous arcs. These arcs are mostly characterized as very sharp and well-defined parabolas, which indicates that they arise from highly anisotropic thin screens (Figure \ref{fig:B1133_dynspec}). In a small number of observations, especially those below 700\,MHz, we see very clear inverted arclets in the secondary spectra (Figure \ref{fig:inverted_arclets}), further confirming that the scattering in this pulsar is highly anisotropic. 

\begin{figure*}%
	\includegraphics[width=14.3cm]{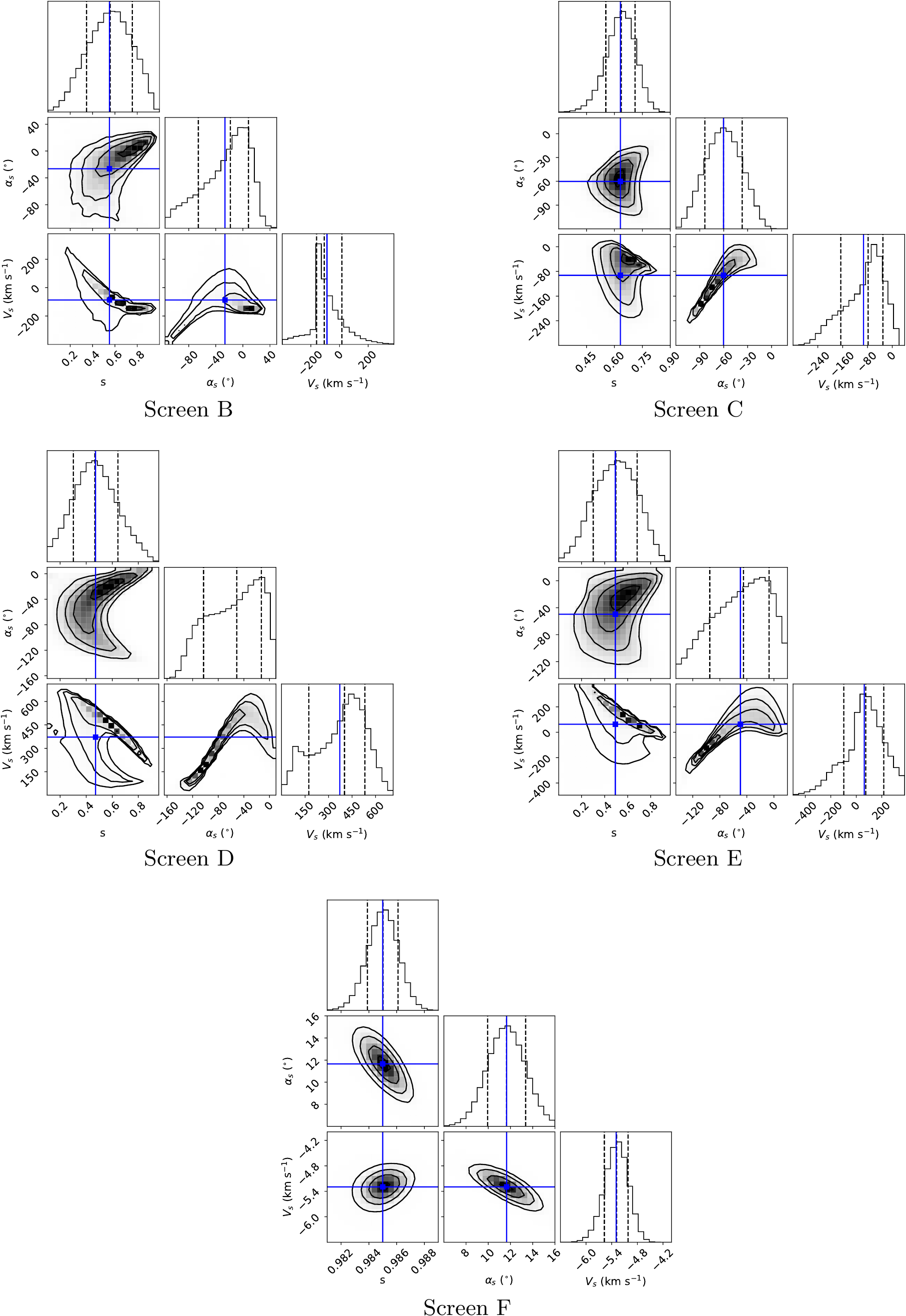}	
	\centering
	\vspace{0.3cm}
    \caption{Corner plots showing the results of MCMC exploration of the parameter space for our screen models. 
    In each case, 300,000 models were generated using flat priors, in a range of $\alpha_{\text{s}}$ corresponding to one specific solution (see text). Dashed lines represent the median, and 0.16, and 0.84 quantiles, while solid lines indicate the distribution mean. While precise screen parameters have been determined for Screens C\ and F, the Screen B, D, and E parameters are not well-constrained.}%
    \label{fig:cornerplots}%
\end{figure*}

\subsection{Arc curvature measurement}\label{sec:curvature}
The high-quality secondary spectra from our observations allow for precise curvature measurements for the scintillation arcs. Our analysis of the secondary spectra arc curvatures made use of the \textsc{scintools}\footnote{https://github.com/danielreardon/scintools} Python module (described in \citealp{rea18,rea20}). This tool set has recently been used in measuring the orbital parameters of PSR\,J0437$-$4715 via the variations in arc curvatures modulated by the pulsar orbit and the Earth orbit \citep{rcb+20}, providing an independent analysis to those of pulsar timing measurements \citep{pdd+19}. It has recently been suggested that particular scattering screen morphologies will cause the apex of the parabolas in the secondary spectrum to be offset from the origin \citep{shi21}. We do not see any evidence for this in our data set, and we assume that the apex of all parabolas is at the origin.

The dynamic spectra were re-sampled in terms of wave numbers rather than frequencies (i.e. $k=2\pi/\lambda$, and see \citealp{fcm+14}) before secondary spectra were computed, to allow arc curvatures from our multi-frequency observations to be directly compared in a consistent manner. This re-sampling also eliminates the wavelength dependence in the scintillation arc curvature, which makes the arcs sharper over a finite bandwidth. The curvature for a scintillation arc in the re-sampled spectra therefore takes the form:
\begin{equation}
    \eta_C = \frac{c}{\lambda^2}\eta = \frac{1}{2} \frac{D_{\rm eff}}{V_{\rm eff}^2\cos^{2}\alpha}~,
\end{equation}
where $\eta$ is the curvature of the scintillation arc without re-sampling, defined in Equation \ref{eqn:eta}.
The scintillation arc curvatures were then measured by an application of a generalized Hough transform (see \citealp{bot+16}). This was done by evaluating a path integral along different trial curvatures, producing a distribution of the integrated power as a function of trial curvature. 
In the resulting distribution the optimum arc curvature and its uncertainty were determined by fitting parabolas to the peaks in integrated power (see the example given shown in Appendix \ref{sec:appendix1}, Figures \ref{fig:hough_transform} and \ref{fig:curvefit_example}). 
Since many of the PSR\,B1133+16 secondary spectra display multiple scintillation arcs (e.g. Figure \ref{fig:B1133_dynspec}), the above process was repeated over several ranges of $\eta$ to avoid identifying only the brightest arc at the global maximum of the integrated power. Final $\eta$ values were checked by over-plotting their resulting parabolas on the secondary spectra, and $\eta$ values with large uncertainties e.g. due to low S/N were excluded. In some cases, anomalously small uncertainties in the curvature measurements were recorded, which appeared to be significantly underestimated. 
To reduce the impact of these measurements on the model fitting (see Section \ref{sec:annual_variations}), we have scaled the uncertainties in quadrature for each arc separately. After a first iteration of model fitting, a factor $E_{\text{quad}}$ was found that produced a reduced $\chi^{2}=1$ when scaling the uncertainties as
$\sigma_{\text{scaled}}=( \sigma_{\text{initial}}^{2}+E_{\text{quad}}^{2}  )^{1/2}$.

We note that several observations appear to have faint scintillation arcs for which we have not been able to precisely measure curvatures and so were not included in our analysis. As the ranges of curvatures that we have measured are consistent for decades (Figure \ref{fig:curvature_vals}, and see below), this implies that all of the arcs are likely to be present in all observations, albeit at a S/N that is too low for us to reliably detect. Taking this to be true, it is not unreasonable to expect that there may be additional arcs that we are not sensitive to, either due to low S/N or being at curvatures that are too high or low to easily distinguish in our observations. A targeted campaign with highly-sensitive next-generation telescopes would potentially uncover many more screens than the six we have presented in this work.

\subsection{Arc curvatures associated with specific screens}\label{sec:arc_curvatures_populations}

When plotting our measured arc curvatures as a function of time we see that several distinct populations emerge.
By noting which arcs appear in secondary spectra with multiple arcs (e.g. Figure \ref{fig:B1133_dynspec}), we are able to unambiguously assign curvatures to particular populations. We show this in Appendix \ref{sec:appendix2} (Figure \ref{fig:multiarcs}), and find that a minimum of six populations are required, which we refer to as Arcs A\,--\,F in ascending order of curvature (Figure \ref{fig:curvature_vals}).
We note that, with the exception of Arcs A and E, all pairs of arcs have been detected in the same observation (see Table \ref{tab:arcdetectiongrid}). Remaining curvatures were assigned to populations based on these ranges, and the annual variation of the curvatures (discussed in detail in Section \ref{sec:annual_variations}).
 Arc C is almost always the brightest arc, and in instances where only one arc is detectable this is always Arc C. There is a large gap between detections of the highest curvatures but we assume the simplest case where they are all part of the same population (Arc F).

As we outlined in Section \ref{sec:theory}, each range of curvatures arises from an individual scattering screen; we have therefore demonstrated there are at least six distinct screens along the $372\pm3$\,pc line of sight to PSR\,B1133+16.

\begin{table}
\caption{Screen parameters obtained from our MCMC model fitting. Due to the small number of detections, a model is not fit for Screen A. There are solutions at integer multiples of $180^{\circ}$ from the listed $\alpha_{\text{s}}$, for which the velocity alternates between positive and negative values. Here we arbitrarily list the solution with the angle closest to zero. We use the astronomer's convention of the angle being measured North through East.}
\label{tab:screenparameters}
\centering 
{\renewcommand{\arraystretch}{1}%
\begin{tabular} {c c c c c}
\hline
\hline
\vspace{0.1cm}
Screen & $s$ & $\alpha_{\text{s}}$ ($^{\circ}$) & $V_{\text{s}}$ (km\,s$^{-1}$) & $D$ (pc) \\
\hline
\hline
\vspace{0.15cm}
B & $0.50^{+0.22}_{-0.23}$ & $-26^{+37}_{-55}$ &
$-101^{+169}_{-98}$ & $186^{+82}_{-87}$ \\
\vspace{0.15cm}
C & $0.637^{+0.074}_{-0.083}$ & $-60^{+23}_{-24}$ &
$-78^{+47}_{-88}$ & $134^{+27}_{-31}$ \\
\vspace{0.15cm}
D & $0.47^{+0.18}_{-0.16}$ & $-52^{+39}_{-51}$ &
$+399^{+135}_{-230}$ & $197^{+67}_{-59}$ \\
\vspace{0.15cm}
E & $0.50^{+0.19}_{-0.20}$ & $-45^{+38}_{-50}$ & $+74^{+142}_{-173}$ & $186^{+69}_{-73}$ \\
F & $0.9853^{+0.0015}_{-0.0016}$ & $+11.3^{+2.4}_{-2.3}$ &
$-5.28^{+0.37}_{-0.42}$ & $5.46^{+0.54}_{-0.59}$ \\
\vspace{-0.3cm}
 & & & & \\
\hline
\hline
\end{tabular}}
\end{table}

\subsection{Annual variation of arc curvatures and determination of screen astrometry}\label{sec:annual_variations}
As mentioned in Section \ref{sec:theory}, the orbital motion of the Earth, together with the proper motion of the pulsar and the scattering screen, leads to annual variations in observed arc curvatures (under the assumption that the scattering geometry remains static throughout a one-year period). As the proper motion and distance of PSR\,B1133+16 are known to high precision, this allows for the astrometric parameters of the scattering screens to be extracted when using the contribution of the Earth's orbit to disentangle the velocities, if the annual signal is well-sampled with curvature measurements of sufficiently-high precision.

When folding all curvature measurements into a single year-long period, the annual signals for Arcs C\,--\,F immediately become apparent. 
For each scintillation arc, we fit the curvature measurements with three parameters: the fractional distance $s$, the transverse velocity $V_{\rm screen}$, and the angle between the major axis of the anisotropic image and the declination axis on the associated scattering screen $\alpha_{\text s}$. As we noted in Section \ref{sec:theory}, only the effective velocity parallel to the major axis can be probed in highly anisotropic scattering. Hence, instead of fitting a vector velocity for the scattering screen, we fit a single scalar along the major axis of the image. Practically, we fit the curvature measurements using the following functional form:
\begin{align}\label{eqn:eta_model}
    &\eta_{\mathrm{C}}(t,s, V_{\rm screen},\alpha_{\text s}) = \nonumber\\
    &\frac{1}{2}\frac{d_{\text{psr}}}{\left(|\textbf{A}(t)|\cos(\alpha_{\rm s}-\alpha_{\rm v}(t))-V_{\text{screen}}/s\right)^{2}}\left(\frac{1-s}{s}\right) ,
\end{align}
where 
\begin{equation}\label{eqn:V_expres}
    \textbf{A}= \textbf{V}_{\text{psr}}\frac{1-s}{s}+\textbf{V}_{\text{Earth}}(t)
\end{equation}
is the known part of the effective velocity,
$t$ is the time of the year, and $\alpha_{\text{v}}(t)$ is the angle between $\textbf{A(t)}$ and the declination axis (so $\alpha_{\rm s}-\alpha_{\rm v}(t)$ is the angle between $\textbf{A(t)}$ and the major axis of the image), which varies due to changing Earth motion.

The screen velocity is canonically expected to be comparable to the speed of sound of interstellar plasma, which is $\sim10$\,km\,s$^{-1}$ \citep{gs95}. However, this is only true in the absence of external factors driving ISM turbulence. Some examples of this are: evaporated material shed from objects in the vicinity of hot stars \citep{wtb+17}, expanding supernova remnants \citep{lr00}, stellar winds, and high-velocity (i.e. supersonic) neutron stars moving in the local ISM and producing shocks. Additionally, material at high galactic latitudes would have an acceleration contribution from their orbital motion in the galactic potential. This means that although screen velocities consistent with the speed of sound might be expected, there are many scenarios that can explain departures from it.

Applying Equation \ref{eqn:eta_model}, we extract the unknown screen parameters: $s$, $\alpha_{\text{s}}$, and $V_{\text{screen}}$. We calculate the pulsar velocity in right ascension ($\alpha$) and declination ($\delta$) using the coordinates, proper motion, and distance obtained from VLBI measurements (presented in \citealp{dgb+19}), as $V_{\alpha}=-129.404^{+0.050}_{-0.018}$\,km\,s$^{-1}$ and $V_{\delta}=642.89^{+0.13}_{-0.10}$\,km\,s$^{-1}$. We made use of the \textsc{Astropy}\footnote{https://www.astropy.org/} package \citep{art+13, aps+18} to calculate the observer motion in the pulsar reference frame, and relative to the Solar System barycenter using the coordinates of the Arecibo Observatory.

\begin{figure}
    \vspace{0.5cm}
	\includegraphics[width=\columnwidth]{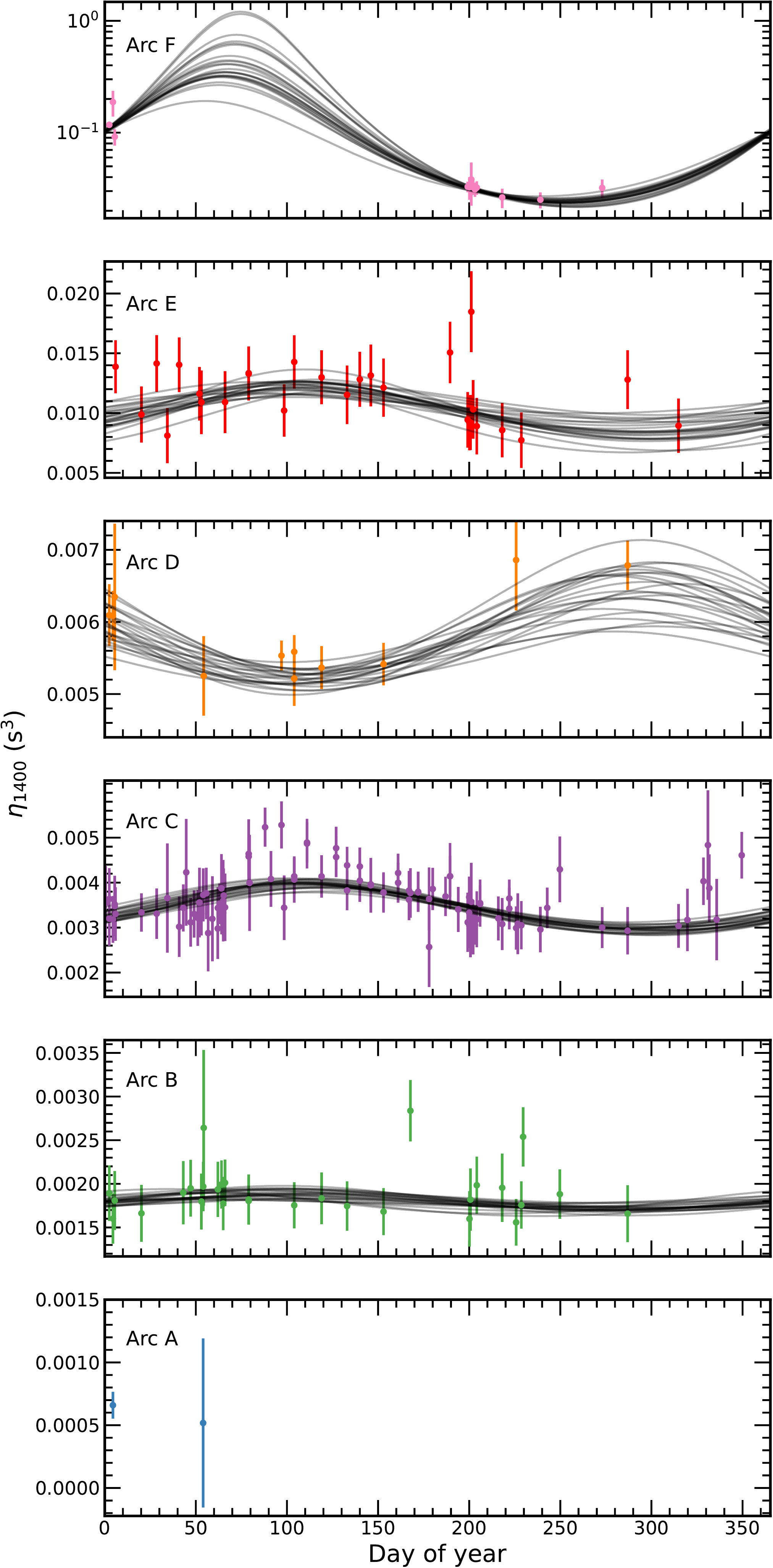}	
	\centering
	\caption{Annual variations in arc curvature $\eta$ for the PSR\,B1133+16 arcs, labeled as Arc A\,--\,F in order of ascending curvature. The curvatures have been scaled to a reference frequency of 1400\,MHz. Black lines show 24 randomly-selected models from the results of our MCMC parameter estimation (Figure \ref{fig:cornerplots}), with the exception of Arc A, due to the small number of detections of this arc. Note that the Arc F plot uses a logarithmic scale due to the high amplitude of the annual variation, while the other arc plots use a linear scale. Our models provide a good description of the periodic variations in Arcs C\,--\,F.} 
	\label{fig:curvature_annual}
\end{figure}

The screen parameters in Equation \ref{eqn:eta_model} are highly covariant, particularly when the amplitude of the periodic curvature variations is smaller than the measurement uncertainties (as is the case for Arcs A and B). We therefore explore the parameter space of the models using a Markov chain Monte Carlo (MCMC) technique. 
As mentioned above, there are values of $\alpha_{\text{s}}$ every $180^{\circ}$ that produce solutions to Equation \ref{eqn:eta_model}.
We chose priors for the model parameters such that only one of the $\alpha_{\text{s}}$ solutions was being tested, which we arbitrarily chose to be the one closest to $0^{\circ}$.
The resulting corner plots from this MCMC analysis are shown in Figure \ref{fig:cornerplots}, and we display the models together with the annual curvatures in Figure \ref{fig:curvature_annual}. The parameters that we have extracted from the fits to the model (Equation \ref{eqn:eta_model}) are listed in Table \ref{tab:screenparameters}. 

We find that our model provides a good description of the observed curvature variations in Arcs C\,--\,F, due to the sufficiently high-precision detections of arcs associated with these screens, compared to the amplitude of their variations in curvature. As mentioned above, we do not attempt to fit a model to Arc A, owing to the few detections we have made of curvatures associated with its screen. Although we have modeled the periodic variations in Arc B, we were not able to find models with a consistent phase, indicating that the amplitude of the annual variation is of comparable or smaller size to our measurement uncertainties. 

Since Arc B is well-sampled throughout the Earth's orbit, it is visible in observations separated by decades (Figure \ref{fig:curvature_vals}), and its range of measured curvatures is relatively low ($\sim0.0014$\,--\,$0.0035\,\text{s}^{3}$, at a reference frequency of 1400\,MHz),
we can be confident that the curvatures we have assigned to this population are indeed associated with a single scattering screen. The annual variations are clearly detectable in Arcs D and E, but the magnitude of the measurement uncertainties prevents precise parameters from being extracted. At our measurement precision, we are only able to very vaguely place Screens B, D, and E at a fractional distance of $\sim0.5$.

In general, screens closer to the Earth give rise to higher amplitudes in the annual variations of arc curvatures, which implies that Screen B is much closer to PSR\,B1133+16 than the upper limit of our distance uncertainty. This also suggests that, at our measurement precision, precise astrometry of Screens A and B (and any that are undiscovered) would require much higher S/N and greater delay and fringe frequency resolutions (i.e. longer observations with higher frequency resolution) than is available with our data set, or coherent phase retrieval (\citealp{wks+08}, \citealp{bbv+21}).
Observations of PSR\,B1133+16 with the new generation of high-gain telescopes will certainly improve the measurement precision of Arcs B, D, and E to a level where precise astrometric values of the underlying screens can be extracted.

We find precise solutions for the parameters of the screens associated with Arcs C and F (Figure \ref{fig:cornerplots}). We see from Figure \ref{fig:curvature_annual} that the models offer an excellent description of our measured curvatures for these two arcs. Although we have relatively few detections of Arc F, the large amplitude of its annual variation, due to its close proximity to the Earth, allows for more precise parameters to be extracted from fits to the data. We note that the amplitude of the annual variation predicted by the best-fit models varies by approximately one order of magnitude (Figure \ref{fig:curvature_annual}), indicating that the precision we have achieved here could be drastically improved, given further measurements throughout the Earth's orbit.

\section{Discussion}\label{sec:discussion}

\subsection{Screen properties}
We have precisely measured the parameters of Screens C and F (Table \ref{tab:screenparameters}).  
The distances we measure are in general agreement with those presented in \cite{ps06a}, although the values in their work used single observations and assumed a screen angle $\alpha_{\text{s}}=0^{\circ}$ i.e. they did not use the knowledge of the orbit to break degeneracies between parameters. Therefore, their reported values represent lower limits on $s$.
Since \cite{ps06a} did not measure temporal variations in arc curvatures, they do not present measurements of the screen angles or velocities.

\begin{figure}
    \vspace{0.5cm}
    \fboxrule=2pt
	\fbox{\includegraphics[width=8.1cm]{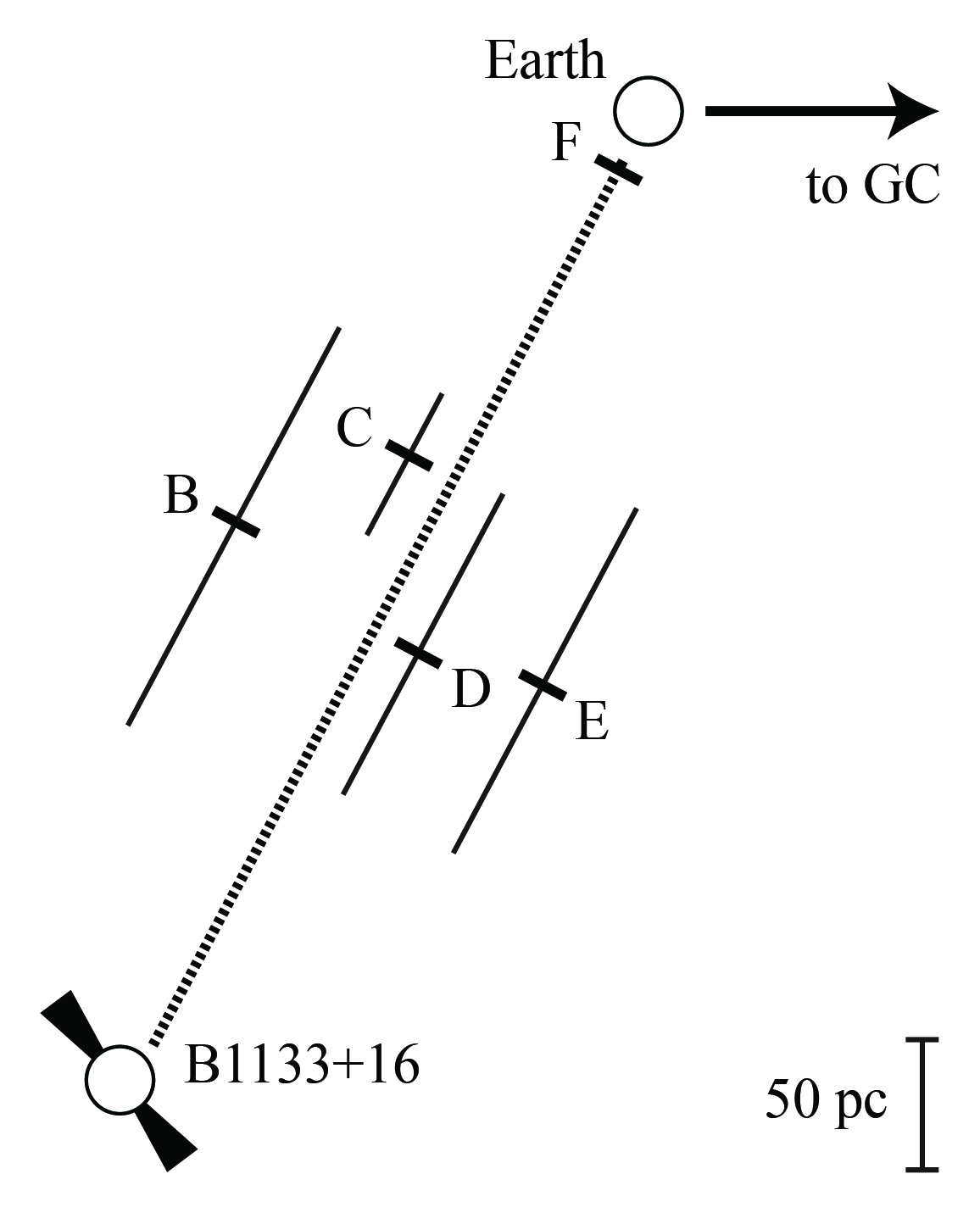}}	
	\centering
	\vspace{0.05cm}
	\caption{Top-down diagram showing the placement of the scattering screens along the line of sight from PSR\,B1133+16 (Galactic longitude $l=241.9^{\circ}$) to the Earth, using the values listed in Table \ref{tab:screenparameters}. The direction to the Galactic center (GC, $l=0^{\circ}$) is displayed. The uncertainty in screen placement is shown by solid lines parallel to the line of sight. Our analysis shows that the observed scattering comes from distinct regions along the line of sight, with the high-curvature arcs arising from screens very local to the Earth. Image credit: Kristi Mickaliger.}
	\label{fig:screens_diagram}
\end{figure}

Although the Solar System is moving at $\sim$ 26\,km\,s$^{-1}$ with respect to the local ISM (e.g. \citealp{wmw15}), and so velocities in the range of approximately $\pm30$\,km\,s$^{-1}$ would be consistent with the ISM speed of sound ($\sim10\,\text{km}\,\text{s}^{-1}$), there appears to be a departure from this in the case of Screen C. While the precision of the screen velocity measurement is quite low, there is a strong preference for a velocity far from that expected from thermal motion of the ISM, which would indicate that the scattering medium has been energized by an external source.

\subsection{The local interstellar medium}\label{sec:localISM}
The close distance of PSR\,B1133+16 combined with the persistence of several discrete scattering screens over decades indicates that the local ISM contains several distinct regions, which we illustrate in Figure \ref{fig:screens_diagram}. It perhaps seems surprising that there are at least six distinct ISM regions within only $372\pm3$\,pc. We do not find evidence for any known structures in the region around PSR\,B1133+16, using the Aladin Sky Atlas\footnote{https://aladin.u-strasbg.fr/aladin.gml} \citep{bf14}, such as the H\,II region that has been identified as the dominant source of the observed scattering in PSR\,B1642$-$03 \citep{grl94} and PSR\,J1643$-$1224 (Mall et al., submitted), nor do we find any reported nearby hot stars.

The lack of associated objects along the line of sight lends weight to the predictions by \cite{rbc87} and \cite{pl14} that thin elongated lenses and corrugated plasma sheets, respectively, formed within the ISM give rise to scattering. However, as mentioned above, we measure a velocity for Screen C which is significantly different from the speed of sound in the ISM that may exclude this explanation. PSR\,B1133+16 has traveled $\sim0.0035$\,pc throughout the duration of our data set, which places a lower limit on the size of the scattering regions we probe.

Remaining explanations are ones where density fluctuations within the local ISM are responsible. The Earth is located in the Local Bubble: an under-dense region of interstellar space driven by the aggregated effects of nearby supernovae over the last $\sim10$\,--\,20\,Myr (e.g. \citealp{fri06}). Maps of the Local Bubble produced by measurements of interstellar absorption \citep{lvv+14} show that the line of sight towards PSR\,B1133+16 cuts through several density boundaries. Although the distance we measure to Screen C is relatively imprecise, it is consistent with the approximate location of an over-density within a cavity of the GSH238+00+09 Super Bubble \citep{hei98}. Screens B, D, and E could possibly be attributed to this region of the Local Bubble as well. While it is possible that Screen F also arises from structure within the Local Bubble, it is extremely nearby, at distances below the resolution of current Local Bubble maps, but unlikely to be described by the same explanation as Screens B\,--\,E.

\subsection{Future prospects}\label{sec:future}
The properties of Screens C and F we have measured here are precise and demonstrate the utility of pulsar scintillation as a means of probing the local ISM. 
We briefly discuss some ways that this work can be built upon to uncover further information about the local ISM.

There are a number of new facilities that have started operations in recent years that will improve the prospects for identifying more pulsars with multiple well-resolved scattering screens, and for making high-precision measurements of arc curvatures. 
The Canadian Hydrogen Intensity Mapping Experiment (CHIME, detailed in \citealp{cab+21}) observes almost every known pulsar in the Northern sky with an approximately daily cadence, in the 400\,--\,800\,MHz band, making it an ideal instrument for measuring temporal variations in ISM properties. 
The Five-Hundred-Metre Aperture Spherical Radio Telescope (FAST, detailed in \citealp{llx19}) has recently been used to measure scattering screens in PSR\,B1929+10 and PSR\,B1842+14, which was not possible with the previous generation of radio telescopes \citep{yzw+20}. 
MeerKAT \citep{bja+20}, the precursor to the Square Kilometre Array (SKA), has recently been utilized in making scattering measurements of a small number of pulsars in the Southern sky \citep{okp+21} and will be further improved with the completion of the full SKA \citep{kea18}.
Independent determination of the screen angle through direct VLBI measurements (in the same way as \citealp{bmg+10}) could potentially offer better constraints on the allowed parameter space of the curvature variation models. This would allow a more precise distance to the screens to be determined.

In this work, we do not detect all six arcs in every one of our observations, and in the case of Arcs A and F, we have very few detections. The results of our analysis tell us that all six arcs are probably present in the secondary spectra. There are likely three reasons for these non-detections: 
\begin{enumerate}
	\item Stochastic variations in the scattering properties of a screen (e.g. as a result of intermittent turbulence, see e.g. \citealp{smc+01}, \citealp{wms+04}, \citealp{crsc06}) may also result in non-detections at a particular epoch. This is expected in the \cite{pl14} corrugated sheet model.
	\item Some of the arcs are simply too faint to be detected in every observation (we note that many different configurations of receivers and backends have been used throughout our data set, and the presence of RFI can quickly diminsh the S/N of secondary spectra arcs).
	\item In the case of Arcs A and F, curvatures at certain parts of the Earth's orbit are too great or small to be identified in our secondary spectra (in Figure \ref{fig:curvature_annual} we see that the predicted amplitude of the Arc F variation is up to an order of magnitude greater than our highest-measured curvature).

\end{enumerate}
Therefore observations with the new generation of highly-sensitive telescopes with broadband recording will allow for both higher precision curvatures to be measured and more consistent detections of the fainter arcs.
PSR\,B1133+16 is special in that six scintillation arcs are now known in its secondary spectra, more than the other pulsars with known multiple scintillation arcs \citep{ps06a}. However, there is seemingly nothing unique about the line of sight to PSR\,B1133+16 to which these multiple scattering screens could be attributed. At the time of writing, this pulsar is the 29$^{\text{th}}$ brightest ($20\pm10$\,mJy at 1400\,MHz) listed in the ATNF pulsar catalogue\footnote{https://www.atnf.csiro.au/research/pulsar/psrcat/} \citep{mhth05}, which makes it likely that it is the intrinsic brightness of the pulsar (and therefore, the scintillation arcs), and the high sensitivity of the Arecibo Telescope, rather than its line of sight which gives rise to a large number of detectable scintillation arcs. It then follows that higher sensitivity observations of PSR\,B1133+16 will uncover previously unknown scintillation arcs and that observations of other pulsars will uncover multiple arcs. Such a census of this pulsar and others will allow for the distinct regions of
the local ISM to be mapped.
As well as providing a robust test of the predictions for the origin of scattering regions by \cite{rbc87}, \cite{pl14}, this will provide a new picture of the way the local ISM is organized, on a distance scale far below that of current local ISM maps.

\section{Conclusions}\label{sec:conclusions}
We have used 34 years of high-sensitivity observations with the Arecibo Telescope to study the scintillation properties of PSR\,B1133+16. By measuring the arc curvatures in the secondary spectrum generated from our observations, we have identified six scintillation arcs that are consistent for decades and which we have attributed to discrete scattering screens. Using the assumption that the properties of the scattering screens have not changed significantly throughout our data set we have modeled the variations in arc curvatures, finding the locations of five of these six scattering screens, and have achieved sub-parsec precision in one case. We speculate that structures within the Local Bubble are responsible for the underlying scattering screens, and suggest that some of the screens may be associated with the boundary of a known region of the Local Bubble, although this is unlikely for the screen closest to the Earth.

\section*{Acknowledgements}
This work is lovingly dedicated to the memory of James McKee's wife, Jeanette Marie Bevan. You were a constant inspiration to me, and I miss you every day my little love. We thank Rik van Lieshout, Marten van Kerkwijk, Fardin Syed, and Ue-Li Pen for valuable discussions, and Kristi Mickaliger for providing the screen placement diagram. We are grateful for the many constructive comments by the anonymous referee, which have greatly improved this paper. J.\,W.\,McKee is a CITA Postdoctoral Fellow: This work was supported by the Natural Sciences and Engineering Research Council of Canada (NSERC), [funding reference \#CITA 490888-16], and by a grant from the US National Science Foundation (2009759; DS and HZ). JMC acknowledges support from the National Science Foundation (NSF AAG-1815242) and is a member of the NANOGrav Physics Frontiers Center, which is supported by the NSF award PHY-1430284. 
The Arecibo Observatory was operated by the University of Central Florida, Ana G. M\`{e}ndez-Universidad Metropolitana, and Yang Enterprises under a cooperative agreement with the National Science Foundation (NSF; AST-1744119). We especially thank the staff of the Arecibo Observatory whose skill, dedication,
and professionalism greatly aided in the collection of the data presented here.

\bibliography{mzsc21}{}

\appendix

\restartappendixnumbering

\section{Curvature measurement via Hough transform}\label{sec:appendix1}
\begin{figure}
    \vspace{0.5cm}
	\includegraphics[width=\textwidth]{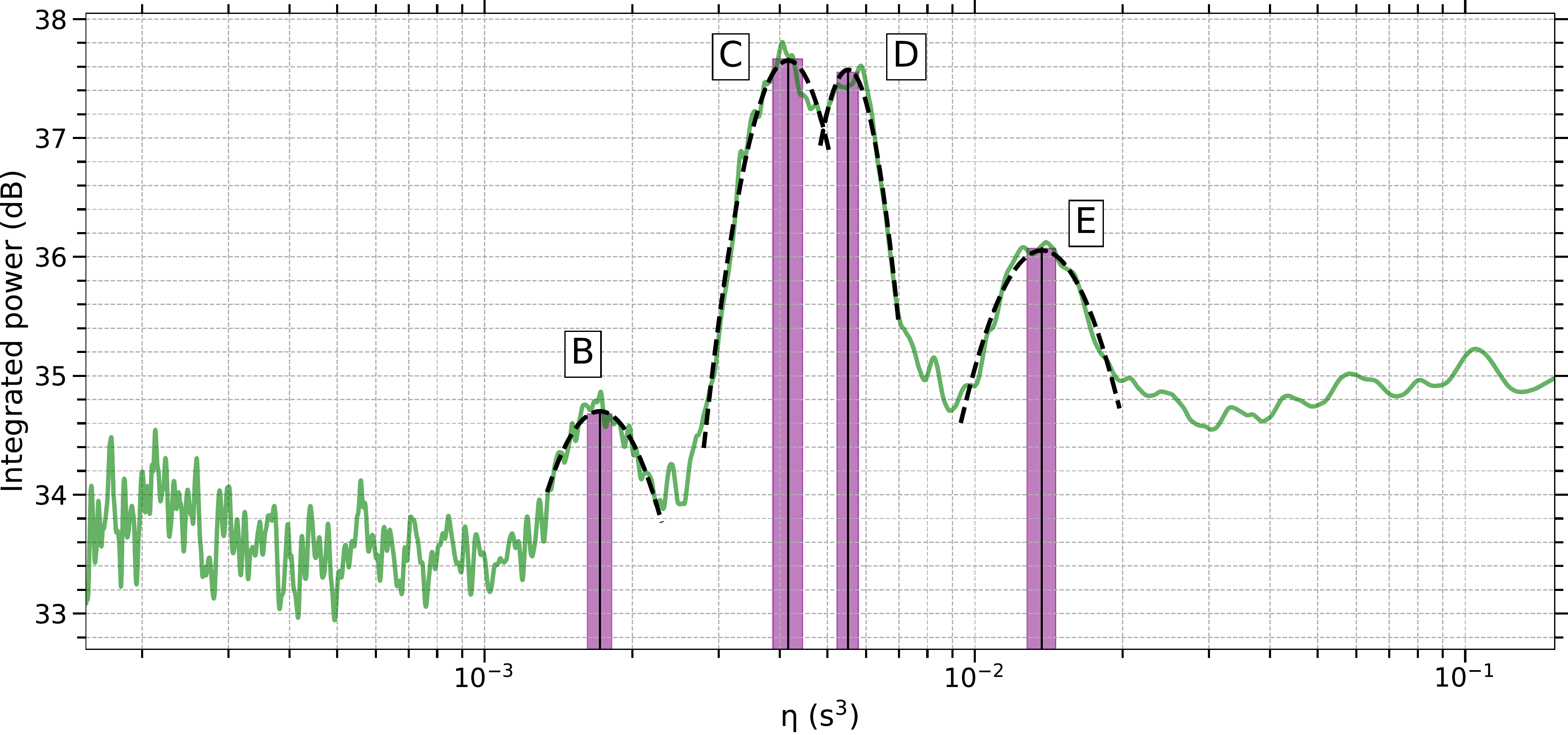}	
	\centering
	\caption{An example of the Hough transform applied to the secondary spectrum of an observation taken at a center frequency of 1450\,MHz on MJD\,57083 (day number 104 in Figure \ref{fig:curvature_annual}). When plotting the integrated power computed for different trial curvatures $\eta$ (green line), four peaks are clearly visible, each corresponding to a resolvable scintillation arc in the secondary spectrum. The best-fit parabolas are over-plotted (black dashed lines), and the corresponding $\eta$ value (solid black lines) and its uncertainty (purple regions) are shown. The measured curvatures are 0.00194(9), 0.0041(1), 0.0057(2), and 0.0136(5) s$^{3}$, and therefore represent Arcs B\,--\,E in order of ascending curvature.}
	\label{fig:hough_transform}
\end{figure}

\begin{figure}
    \vspace{0.5cm}
	\includegraphics[width=8cm]{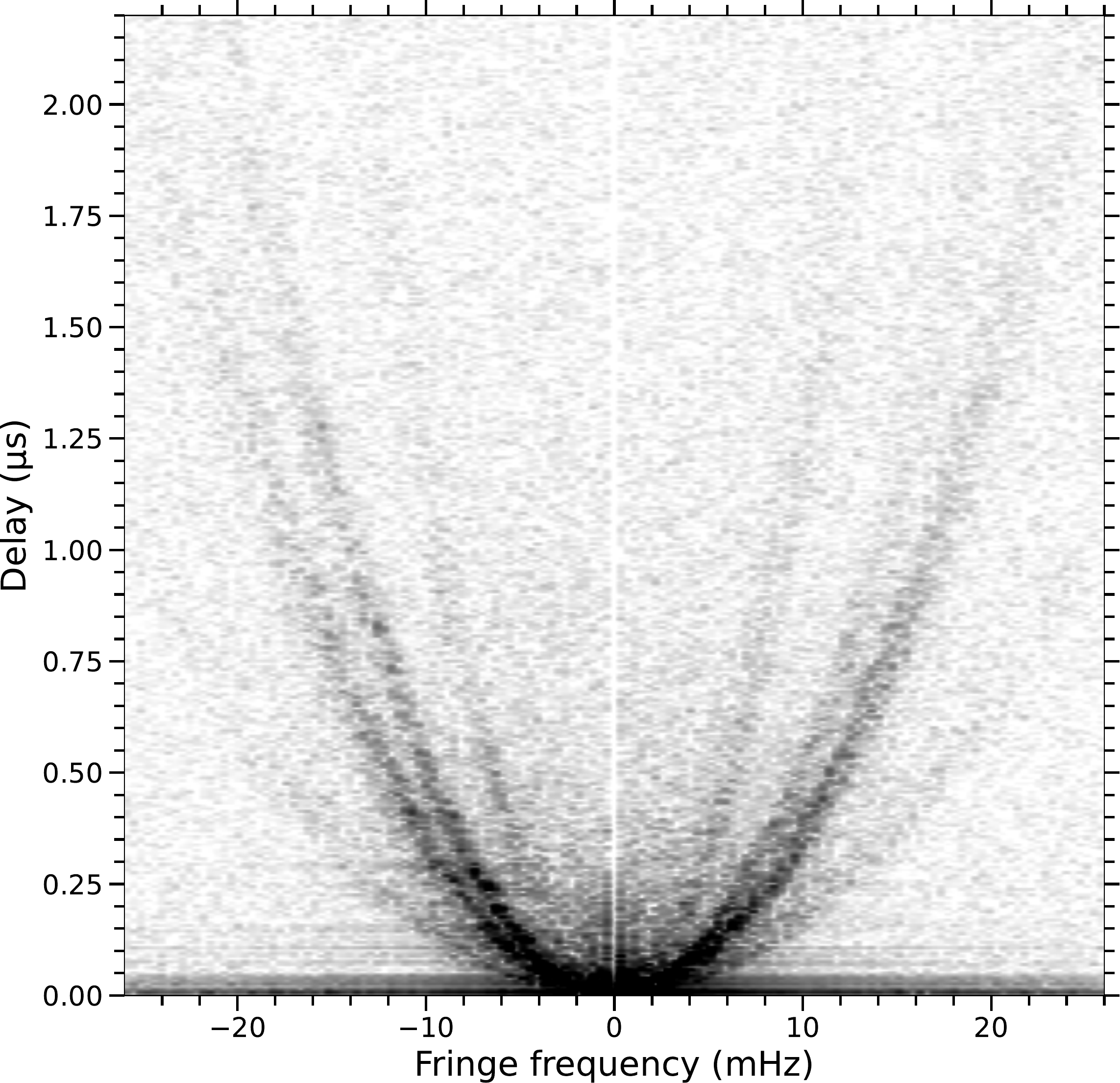}
	\includegraphics[width=7.28cm]{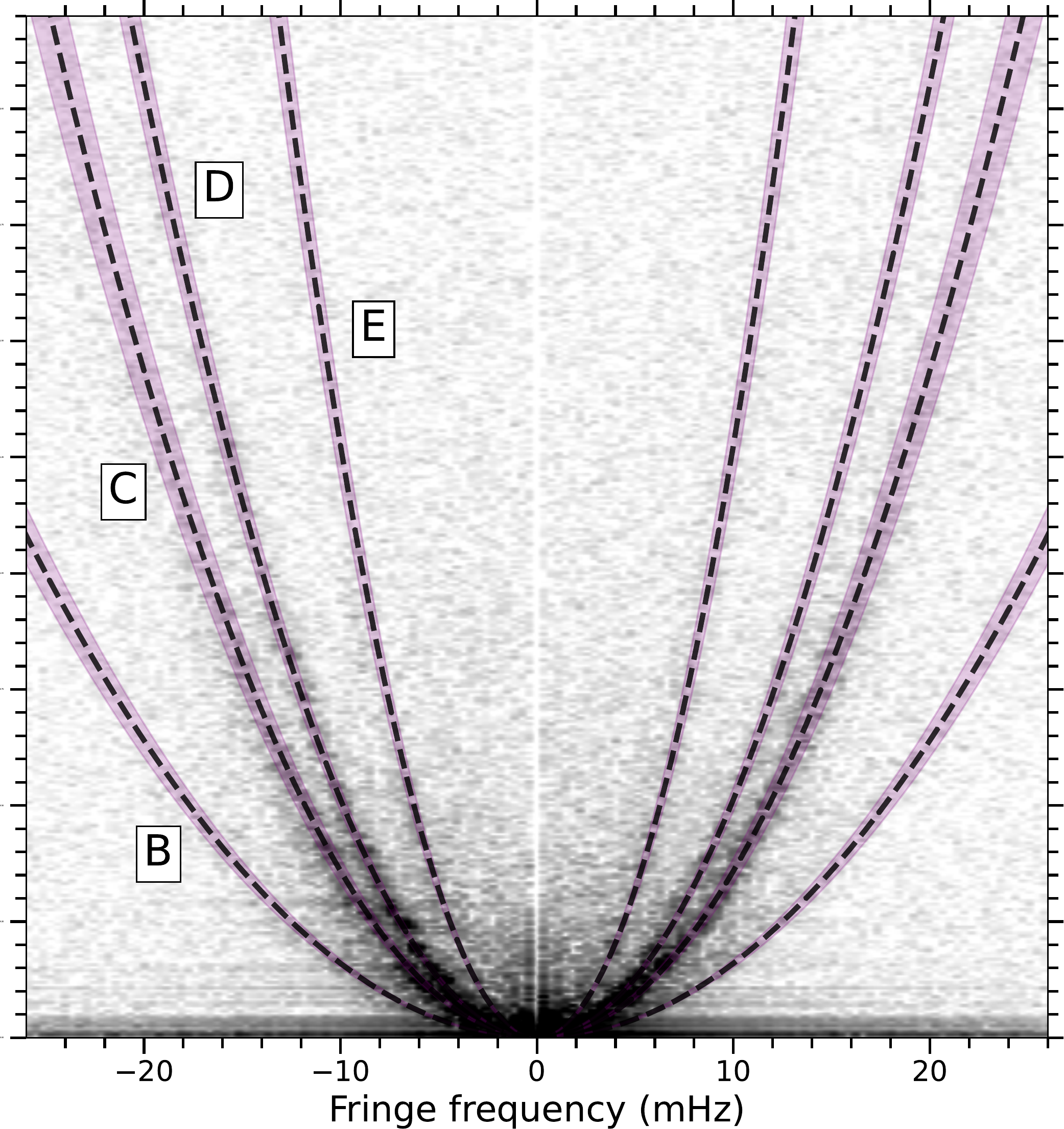}	
	\centering
	\caption{The secondary spectrum used in the Hough transform example above (Figure \ref{fig:hough_transform}). Four scintillation arcs, arising from Screens B\,--\,E, are clearly visible. In the right-hand panel, we overplot the best-fit arc curvatures (black dashed lines) and their uncertainties (purple regions), measured via the Hough transform, and see that our measured values offer an excellent description of the scintillation arc curvatures observed in the secondary spectrum.}
	\label{fig:curvefit_example}
\end{figure}

In this section, we show an example of how curvatures were measured from our secondary spectra through a generalized Hough transform. See Figures \ref{fig:hough_transform} and \ref{fig:curvefit_example} for details of the method.


\section{Assigning curvatures to populations based on multiple detections of arcs}\label{sec:appendix2}

\begin{figure}
    \vspace{0.5cm}
	\includegraphics[width=17cm]{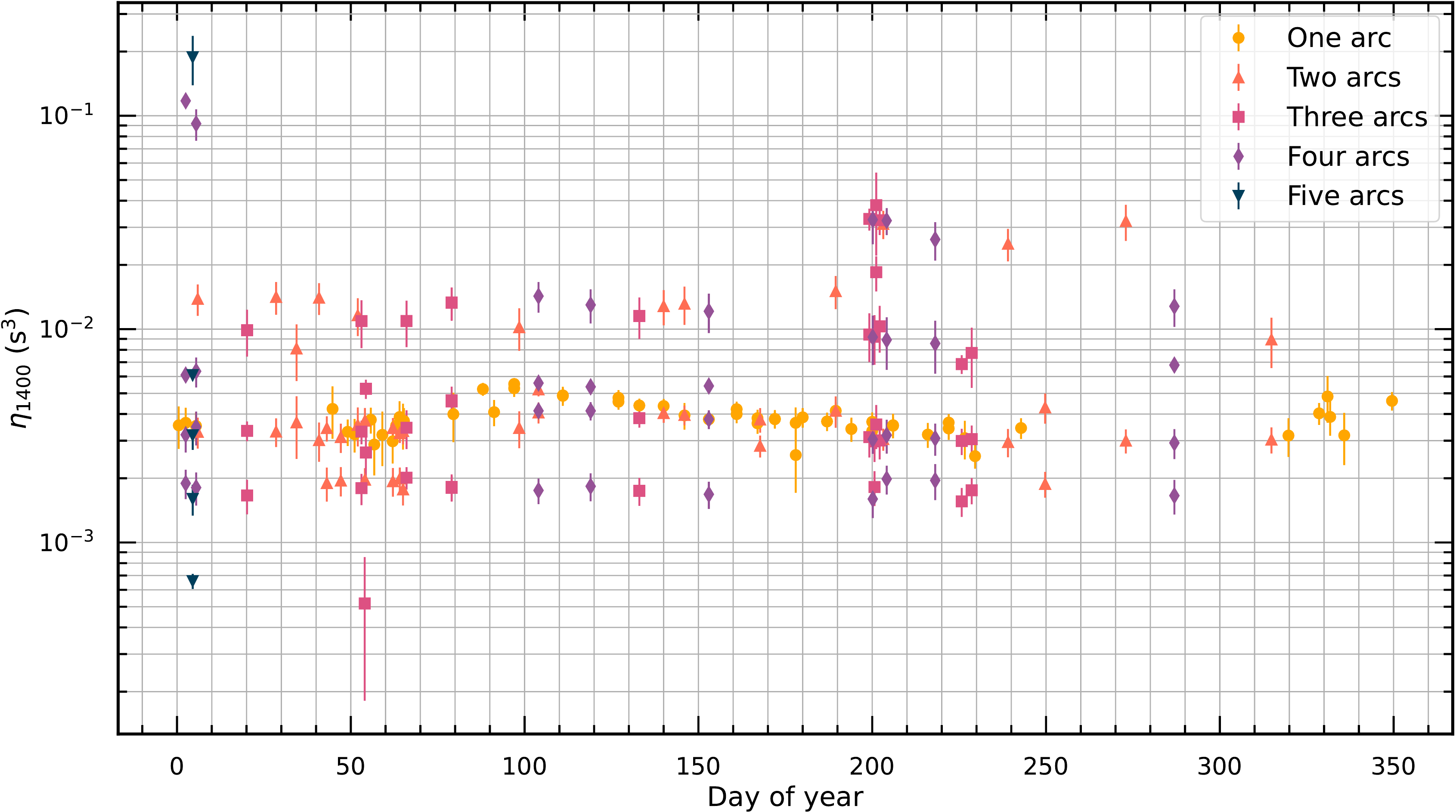}	
	\centering
	\caption{Arc curvatures color coded by the number of arcs detected in that observation. This allows us to unambiguously assign curvatures to a particular population.}
	\label{fig:multiarcs}
\end{figure}

We assigned curvatures to different populations by checking observations with multiple arc detections. We show this in Figure \ref{fig:multiarcs}, and list the number of detections of pairs of screens in Table \ref{tab:arcdetectiongrid}.

\begin{table}
\caption{Number of simultaneous detections of Screens A\,--\,F in our data set.}
\label{tab:arcdetectiongrid}
\centering 
{\renewcommand{\arraystretch}{1}%
\begin{tabular} {| c | c | c | c | c | c | c |}
\hline
\vspace{0.1cm}
\textbf{Screen} & \textbf{A} & \textbf{B} & \textbf{C} & \textbf{D} & \textbf{E} & \textbf{F} \\
\hline
\textbf{A} & - & 2 & 1 & 2 & 0 & 1 \\
\hline
\textbf{B} & 2 & - & 27 & 9 & 15 & 6 \\
\hline
\textbf{C} & 1 & 27 & - & 9 & 28 & 12 \\
\hline
\textbf{D} & 2 & 9 & 9 & - & 4 & 3 \\
\hline
\textbf{E} & 0 & 15 & 28 & 4 & - & 6 \\
\hline
\textbf{F} & 1 & 6 & 12 & 3 & 6 & - \\
\hline
\end{tabular}}
\end{table}

\end{document}